
\input panda
\loadamsmath
\pageno=0
\baselineskip=18pt
\nopagenumbers{
\line{\hfill CERN-TH/95-63}
\line{\hfill DFTT 19/95 }
\ifdoublepage \bjump\bjump\bjump\bjump\else\vfill\fi
\centerline{\bf Cosmological Solutions of Higher-Curvature String
  Effective Theories }
\centerline{\bf with Dilatons}
\bjump
\centerline{M. C. Bento\footnote{$^*$}{On leave of absence from Departamento
de F\' \i sica, Instituto
 Superior T\'ecnico, Av. Rovisco Pais, 1096 Lisboa Codex, Portugal}}
\centerline{\it CERN, Theory Division}
\centerline{\it CH-1211  Geneva 23}
\centerline{\it Switzerland}
\vskip 0.2cm
\centerline{ and}
\vskip 0.2cm
\centerline{O. Bertolami$^*$\footnote{$^\dagger$}{Also at Theory Division,
CERN.}}
\centerline{INFN-\it Sezione Torino}
\centerline{\it Via Pietro Giuria 1}
\centerline{\it I-10125  Turin, Italy}
\bjump
\ifdoublepage
\vfill
\noindent
\line{CERN-TH/95-63\hfill}
\line{DFTT 19/95 \hfill}
\line{March 1995\hfill}
\eject\null\vfill\fi
\centerline{\bf Abstract}\sjump

We study the effect of higher-curvature terms in the string
low-energy effective actions on the cosmological solutions of the theory, up to
 corrections quartic in the curvatures, for the bosonic and  heterotic
strings as well as for the
type II superstring. We find that  cosmological solutions exist for all
string types but they  always  disappear  when the dilaton field is included, a
conclusion that can be avoided if string-loop effects are taken into account.

\sjump\vfill
\ifdoublepage\else
\noindent
\line{CERN-TH/95-63\hfill}
\line{DFTT 19/95 \hfill}
\line{March 1995\hfill}\fi
\eject}
\yespagenumbers\pageno=1

\def\ni{\noindent}
\def\laq{\raise 0.4ex\hbox{$<$}\kern -0.8em\lower 0.62 ex\hbox{$\sim$}}
\def\gaq{\raise 0.4ex\hbox{$>$}\kern -0.7em\lower 0.62 ex\hbox{$\sim$}}


The fact that string theories, in their low-energy limit, give rise to
effective theories of gravity containing higher-curvature corrections
to the usual scalar curvature term, has led to renewed interest in
such extensions of Einstein gravity [\Ref{bunch} -- \Ref{bento}]. The
leading quadratic correction, proportional to the Gauss-Bonnet
combination, in particular, has ben studied in some detail in
[\Ref{desera}], where it is shown that there are cosmological (de
  Sitter) solutions in addition to Minkowski space and that they are
  unstable.   For the bosonic string, the next-to-leading corrections
are cubic in the curvatures and it has ben shown that cosmological
solutions remain unstable at this order [\Ref{bento}]. It is, however,
possible that higher curvature terms could
give rise to stable de Sitter solutions.

 In this article, we study the
effect of higher order corrections  to string low-energy
effective actions, up to the quartic curvature level,
on the existence and stability of the cosmological solutions
to these theories.  For the type II superstring
there are neither quadratic nor cubic-curvature corrections
whereas for the heterotic
string only the  cubic terms are absent  so that the
quartic curvatures are actually the leading
and next-to-leading corrections to these theories, respectively.
We shall study also the
consequences of the inclusion of the dilaton in the effective action;
for the quadratic correction, it is known that
the inclusion of the  dilaton has dramatic
consequences for the cosmological branches, which become excluded at
this level [\Ref{deserb}]. Finally, we shall analyse the effect of
introducing string-loop corrections to the string effective action.

We start with the action in the so-called s-parametrisation:

$$S= \int d^D x \sqrt{-g}\left[{R\over k} + c_1\alpha^\prime e^{- 2 \phi}
{\cal L}_2 + c_2 \alpha^{\prime 2} e^{- 4 \phi}{\cal L}_3 + c_3
\alpha^{\prime 3} e^{- 6 \phi}{\cal L}_4 + ...\right]
\nfr{a}
where  $k$ is Newton's constant
in D dimensions, $\alpha^\prime $ is the string expansion parameter
and the dots denote terms involving derivatives of the
dilaton which will not be relevant for our discussion.
Here and throughout the article we use the signature $(-,
+,... ,+)$ and the conventions:
$R^\mu_{\phantom{\mu}\nu\alpha\beta}=\Gamma^\mu_{\nu\beta,\alpha} - ...,\
R_{\mu\nu}=R^\lambda_{\phantom{\lambda}\mu\lambda\nu}$. The
$ {\cal L}_i$ terms involve appropriate powers of the curvature, which
have been found by comparison of the
amplitudes generated by the action  \a\ with the string
amplitudes (equivalent results have been obtained  through the computation
of the relevant
$\sigma$-model $\beta$-functions), as follows  [\Ref{bunch2}]

$$\eqalignno{{\cal L}_2 &=\Omega_2, & \nameali{ba}\cr
             {\cal L}_3 &= 2\ \Omega_3 +
R^{\mu\nu}_{\phantom{\alpha\beta}\alpha \beta}
R^{\alpha\beta}_{\phantom{\lambda\rho}\lambda\rho}
R^{\lambda\rho}_{\phantom{\mu\nu}\mu\nu}, & \nameali{bb}\cr
                 {\cal L}_4 &=\zeta(3)
          \left[R_{\mu\nu\rho\sigma}R^{\alpha\nu\rho\beta}\left(
                R^{\mu\gamma}_{\phantom{\mu\gamma}\delta\beta}
                R_{\alpha\gamma}^{\phantom{\delta\sigma}\delta\sigma}
           - 2  R^{\mu\gamma}_{\phantom{\mu\gamma}\delta\alpha}
                R_{\beta\gamma}^{\phantom{\beta\gamma}\delta\sigma}
                                               \right)\right]&   \cr
&\phantom{=}-\xi_H\Biggl[{1\over 8} \left(
              R_{\mu\nu\alpha\beta} R^{\mu\nu\alpha\beta}\right)^2
 +{1\over 4}  R_{\mu\nu}^{\phantom{\gamma\delta}\gamma\delta}
              R_{\gamma\delta}^{\phantom{\rho\sigma}\rho\sigma}
              R_{\rho\sigma}^{\phantom{\alpha\beta}\alpha\beta}
              R_{\alpha\beta}^{\phantom{\mu\nu}\mu\nu}&  \cr
& \phantom{=-\xi_H\Biggl[} -
{1\over 2} R_{\mu\nu}^{\phantom{\alpha\beta}\alpha\beta}
           R_{\alpha\beta}^{\phantom{\rho\sigma}\rho\sigma}
           R^\mu_{\phantom{\mu}\sigma\gamma\delta}
           R_\rho^{\phantom{\rho}\nu\gamma\delta}
         - R_{\mu\nu}^{\phantom{\gamma\delta}\alpha\beta}
           R_{\alpha\beta}^{\phantom{\rho\sigma}\rho\nu}
           R_{\rho\sigma}^{\phantom{\alpha\beta}\gamma\delta}
R_{\gamma\delta}^{\phantom{\mu\nu}\mu\sigma}\Biggr]&   \cr
&\phantom{=} - {1\over 2}\xi_B \biggl[ \left(
                R_{\mu\nu\alpha\beta}R^{\mu\nu\alpha\beta}\right)^2
           - 10  R_{\mu\nu\alpha\beta}
              R^{\mu\nu\alpha\sigma}
              R_{\sigma\gamma\delta\rho}
              R^{\beta\gamma\delta\rho}& \cr
& \phantom{= - xi_B \biggl[}
          -  R_{\mu\nu\alpha\beta}
              R^{\mu\nu\rho}_{\phantom{\rho\sigma\delta}\sigma}
              R^{\beta\sigma\gamma\delta}
              R_{\delta\gamma\rho}^{\phantom{\sigma\beta\rho}\alpha}\biggr]
                                              , &\nameali{bc}\cr}$$
where $\xi_{H(B)}=1$ for the heterotic (bosonic) string and vanishes
for the other string
types  and $\Omega_2,\ \Omega_3$ are the second and third
order Euler densities:

$$\eqalignno{\Omega_2 &=R_{\mu\nu\alpha\beta}^2 -4 R_{\mu\nu}^2 + R^2,
&\nameali{ca}\cr
             \Omega_3 & =R^{\mu\nu}_{\phantom{\mu\nu}\alpha \beta}
R^{\alpha\beta}_{\phantom{\lambda\rho}\lambda\rho}R^{\lambda\rho}_{\phantom{\lambda\rho}\mu\nu}
-2R^{\mu\nu}_{\phantom{\lambda\rho}\alpha\beta}R_\nu^{\phantom{\nu}\lambda\beta\rho}R^\alpha_{\phantom{\alpha}\rho\mu\lambda}+{3\over
4} R R_{\mu\nu\alpha\beta}^2 & \cr
                     &\phantom{=6} + 6  R^{\mu\nu\alpha\beta}
R_{\alpha\mu}R_{\beta\nu} + 4
R^{\mu\nu}R_{\nu\alpha}R^\alpha_{\phantom{\alpha}\mu} - 6
R R_{\alpha\beta}^2 + {1\over 4} R^3. &\nameali{cb}\cr}$$
\noindent
 Coefficients $(c_1,\ c_2,\ c_3)$  are different for  different
string theories

$$\eqalignno{&bosonic :\ \left({1\over 4}, {1\over 48}, {1\over 8}\right),
&\nameali{da}\cr
             &heterotic :\ \left({1\over 8}, 0, {1\over 8}\right),
&\nameali{db}\cr
             &superstring II :\ \left(0, 0, {1\over 8}\right).
&\nameali{dc}\cr}$$

\noindent
For maximally symmetric spaces

$$ R^\mu_{\phantom{\mu} \nu\lambda\sigma}=\Lambda\left(
\delta^\mu_\lambda g_{\nu\sigma} - \delta^\mu_\sigma g_{\nu\lambda}\right),
\nfr{e}
and the dilaton is constant. Hence, the gravitational and scalar field
equations reduce to, respectively

$$\eqalignno{ f(\Lambda)\equiv\Lambda(c_3 \rho \Lambda^3 +  c_2 \sigma
\Lambda^2 + c_1 \beta  \Lambda  + \alpha) & = 0, &\nameali{fa}\cr
g(\Lambda)\equiv \Lambda^2 (3 c_3\rho  \Lambda^2 + 2 c_2\sigma \Lambda +
c_1 \beta ) & = 0,&
\nameali{fb}\cr}$$
where

$$\eqalignno{\alfa &= (D-2) (D-1)   ,&\nameali{ga}\cr
              \beta &=\alpha^\prime k e^{- 2 \phi} (D-4) (D-3)(D-2) (D-1),
     &\nameali{gb}\cr
               \sigma &=  (\alpha^{\prime}k)^2  e^{- 4 \phi}2
                (D-6)(D-1)[2 +  (D-5) (D-4)(D-3) (D-2)] ,&\nameali{gc}\cr
              \rho &=(\alpha^{\prime}k)^3  e^{- 6 \phi} (D-1)(D-2)
          \bigl[- 3 \zeta(3) (D-3) + \xi_H (D-2)(D-9)& \cr
            &\phantom{= \alpha^{\prime 3} e^{- 6 \phi}](D-1)(D-2)\biggl[}
             - 2 \xi_B [(D-2)(D-11) - 1]\bigr] ,&\nameali{gd}\cr}$$

In the absence of the dilaton, only eq. \fa\ has to be satisfied; clearly,
Minkowski space is always a solution and, with our conventions,
de Sitter (anti-de Sitter)
solutions correspond to the real positive (negative) roots of \fa. We
find that, for the
superstring II in the critical number of dimensions $(D = 10)$, there
is only one positive real root, $\Lambda_s=\left({-\alpha\over c_3
\rho}\right)^{1/3}$. For the
heterotic string, there is only one positive real root in $D=4$  and
three real roots in $D=10$ (two negative and one positive).
For the bosonic
string $(D=26)$ there is only one positive real root.

Regarding the stability of these solutions for small graviton excitations about
the background, notice that it can be deduced immediately from the
observation that the sign of $\delta^2 S$ is determined by the sign of
$f^\prime(\Lambda)$ at the roots [\Ref{desera}]; in this way,
we have checked that all  cosmological branches are  unstable except
for one of the anti-de Sitter branches of the heterotic string in $D=10$.

When the dilaton is included, eq. \fb\ has to be taken into account as
well; this is a strong constraint and, indeed, it is easy to check
that the  cosmological  branches we found above do not survive and
only Minkowsky space remains a solution.
We shall see in the following that this conclusion can be evaded  if
string-loop effects are taken into
account. In the so called $\sigma$-parametrization
of the effective action, namely

$$S^{(\sigma)}= \int d^D x \sqrt{-g}e^{-2\phi}\left[{R\over k} +
c_1\alpha^\prime
{\cal L}_2 + c_2 \alpha^{\prime 2} {\cal L}_3 + c_3
\alpha^{\prime 3}{\cal L}_4 + ...\right],
\nfr{j}
these can be included in a rather simple way, i.e. substituting the
exponential $ e^{-2 \phi}$ by the series in the string coupling $g_S \equiv e^
{2 \phi}$ [\Ref{Damour}]:

 $$ B(\phi)\equiv e^{-2\phi}+ a_0 + a_1 e^{2 \phi} + a_2 e^{4\phi} + ...   ,
\nfr{l}
where the coefficients $a_0$, $a_1$, ... are, at the present level of
understanding of string theory, unknown.

The s-parametrisation of action  \a,  is
obtained from the $\sigma$-parametrization, eq. \j, through the conformal
transformation $ g_{\mu\nu}\rightarrow e^{- 2 \phi} g_{\mu\nu}$, or,
if string-loop effects are included:

$$ g_{\mu\nu}\rightarrow B(\phi) g_{\mu\nu},
\nfr{m}
which, in turn, translates into the substitutions

$$ e^{-2 n \phi} \rightarrow  B^n(\phi), \qquad n=1,2,3, ...
\nfr{n}
in \a. As a consequence, eqs. \fa\ and \fb\ now become

$$\eqalignno{\bar f(\bar\Lambda)\equiv\bar \Lambda(\bar \rho c_3{\bar
\Lambda}^3 +\bar
\sigma c_2{\bar \Lambda}^2 + \bar \beta  c_1
\bar \Lambda  + \alfa) & = 0, &\nameali{oa}\cr
\bar g(\bar \Lambda)\equiv {B^\prime(\phi)\over B(\phi)} {\bar \Lambda}^2 (3
\bar\rho
c_3 {\bar \Lambda}^2 + 2
\bar \sigma c_2 \bar \Lambda +\bar \beta c_1) & = 0,&
\nameali{ob}\cr}$$
where $\bar \rho = B^3(\phi)  e^{6 \phi}\rho$,
$\bar \sigma =  B^2(\phi) e^{4 \phi}\sigma$ and $\bar \beta = B(\phi)
e^{2\phi}  \beta$. Notice now the
appearence of the factor $B^\prime(\phi)$ in front of the dilaton equation
of motion; this allows us to require that this factor vanishes in
order to satisfy eq. \ob, which fixes the value of the dilaton field,
$\phi=\phi_0$,
so that there is only one algebraic equation for $\bar \Lambda$. The
analysis then proceeds as for the solutions to eq. \fa, except that
now results  depend  on the value  of
$B(\phi_0)$. For instance, for the
superstring II, where
 $\bar\Lambda_s=B^{-1}(\phi_0)\Lambda_s$ and
$f^\prime(\bar\Lambda_s)=-3\alpha$,
 so that  the sign of the solution indeed
depends on $B(\phi_0)$ but  the stability analysis
does not  change as compared to the case with no dilaton. The
same result holds for the heterotic string in $D=4$.

We stress that our conclusions regarding the existence and stability
of cosmological solutions to higher-curvature string effective theories,
upon inclusion
of next-to-leading  corrections, support and  reinforce the ones that were
already pointed out at the leading-order level, namely that there exist
 de Sitter
branches  that are, however, erased in the presence of the dilaton; given
that this is basically due to the fact that the dilaton equation of motion
represents an  additional strong constraint on possible maximally
symmetric solutions, a feature that remains   when higher
order corrections are
included,  it is therefore natural  to conjecture that this may be  a rather
general result.  Our analysis shows, however,
that this conclusion is not unescapable if  string-loop corrections are
taken into account.

Thus, our results show that string theory has no intrinsic cosmological
problem, at least in $D=4$. Furthermore, it is rather easy to show that the
inclusion of an
explicit cosmological constant may change considerably our conclusions as
stable cosmological branches can be then found. This implies, for instance,
that in order to achieve a sufficiently long lasting period of inflation a
potential for the dilaton with at least one single local minimum is
required. This indicates the prominance of the issue of
supersymmetry breaking and the ensued cosmological constant problem in any
realistic stringy cosmological scenario.

There is, nonethess, a basic ambiguity in our analysis,
associated with the fact that the form we have chosen for the
higher-curvature corrections, eqs. \ba --\bc, is just one among various
possible
parametrizations  that are related by metric
redefinitions  of the type
$g_{\mu\nu} \rightarrow g_{\mu\nu}+\alpha^\prime(a_1 R_{\mu\nu} + a_2
g_{\mu\nu} R) + \alpha^{\prime 2}
(R_{\mu\alpha\beta\gamma}R_{\nu}^{\phantom{\nu}\alpha\beta\gamma} +...) + ...$,
which do not change the S-matrix. In fact, it is possible to show
that there are both  ambiguous and fixed terms according to  whether they are
 changed by field redefinitions or not.
At order $\alpha^\prime$, only the $R_{\mu\nu\rho\sigma}^2$ term is
fixed by the string amplitudes and the Gauss-Bonnet form is singled
out only if we introduce the further requirement that the effective theory be
unitary at this level [\Ref{Zwieb}]. However, it has been shown that at
cubic and  higher orders, it is not possible to write
the effective action only in terms of Euler densities as required
by unitarity. The ambiguous terms are then often chosen so
as to keep the effective action as simple as possible at each order; however,
 even this
choice is not unique since such terms cannot,
in general, be all set to zero simultaneously given that there are relations
among their variations.
We have studied other possible forms for the effective action at the
cubic curvature level, where a complete field redefinition analysis
has been performed in Refs. [\Ref{bento2}, \Ref{bento3}], and we find  that
our conclusions do not basically change, i.e. although there indeed
exist different cosmological solutions for different actions, they are
still always erased in the presence of  the dilaton  and saved by the
inclusion of string-loop effects.

\ni
{\bf Acknowledgements}

\ni
We would like to thank Prof. V. de Alfaro for discussions on the subject of
this paper.

\references

\beginref

\Rref{bunch} {J. Madore, Phys. Lett. A110 (1985) 289; A111 (1985) 283;
F. M\"uller-Hoissen, Phys. Lett. B163 (1985) 106;
J.T. Wheeler, Nucl. Phys. B268 (1986) 36;
D.L. Wiltshire, Phys. Lett. B169 (1986) 36.}

\Rref{desera} {D.G. Boulware and S. Deser, Phys. Rev. Lett. 55 (1985) 2656.}

\Rref{bento} {M.C. Bento and O. Bertolami, Phys. Lett. B228 (1989) 348.}

\Rref{deserb} {D.G. Boulware and S. Deser, Phys. Lett. B175 (1986) 409.}

\Rref{bunch2} {D.G. Gross and E. Witten, Nucl. Phys. B277 (1986) 1;
Y. Kikuchi, C. Marzban and Y.J. Ng, Phys. Lett. B176 (1986) 57;
M.T. Grisaru and D. Zanon, Phys.Lett. B277 (1986) 199;
Q.-H. Park and D. Zanon, Phys. Rev. D35 (1987) 4038;
R.R. Metsaev and A.A. Tseytlin, Phys. Lett. B185 (1987) 52;
 I. Jack, D.R.T. Jones and N. Mohammedi, Phys. Lett. B220 (1989) 171.}

\Rref{Zwieb}{B. Zwiebach, Phys. Lett. B156 (1985) 315.}

\Rref{Damour} {T. Damour and A.M. Polyakov, Nucl. Phys. B423 (1994) 532.}

\Rref{bento2} {M.C. Bento, O. Bertolami, A.B. Henriques and J.C. Rom\~ao, Phys.
Lett. B218 (1989) 162; B220 (1989) 113.}

\Rref{bento3} {M.C. Bento, O. Bertolami and
J.C. Rom\~ao, Phys. Lett. B252 (1990) 401; Int. J. Mod. Phys. A6 (1991) 5099.}

\endref

\ciao

%
%
%
%
%
%
%
%
\def\standardrisposta{s }\def\reducedrisposta{r }
\def\mplarisposta{mpla }\def\zerorisposta{z }
\def\doublerisposta{d }\def\cartarisposta{e }\def\amsrisposta{y }
\newcount\ingrandimento \newcount\sinnota \newcount\dimnota
\newcount\unoduecol \newdimen\collhsize \newdimen\tothsize
\newdimen\fullhsize \newcount\controllorisposta \sinnota=1
\newskip\infralinea  \global\controllorisposta=0
\immediate\write16 { ********  Welcome to PANDA macros (Plain TeX,
AP, 1991) ******** }
\immediate\write16 { You'll have to answer a few questions in
lowercase.}
\message{>  Do you want it in double-page (d), reduced (r)
or standard format (s) ? }\read-1 to\risposta
\message{>  Do you want it in USA A4 (u) or EUROPEAN A4
(e) paper size ? }\read-1 to\srisposta
%
%
%
\def\srisposta{e }
\def\arisposta{y }
\ifx\risposta\standardrisposta \ingrandimento=1200
\message {>> This will come out UNREDUCED << }
\dimnota=2 \unoduecol=1 \global\controllorisposta=1 \fi
\ifx\risposta\reducedrisposta \ingrandimento=1095 \dimnota=1
\unoduecol=1  \global\controllorisposta=1
\message {>> This will come out REDUCED << } \fi
\ifx\risposta\doublerisposta \ingrandimento=1000 \dimnota=2
\unoduecol=2   \message {>> You must print this in
LANDSCAPE orientation << } \global\controllorisposta=1 \fi
\ifx\risposta\mplarisposta \ingrandimento=1000 \dimnota=1
\message {>> Mod. Phys. Lett. A format << }
\unoduecol=1 \global\controllorisposta=1 \fi
\ifx\risposta\zerorisposta \ingrandimento=1000 \dimnota=2
\message {>> Zero Magnification format << }
\unoduecol=1 \global\controllorisposta=1 \fi
\ifnum\controllorisposta=0  \ingrandimento=1200
\message {>>> ERROR IN INPUT, I ASSUME STANDARD
UNREDUCED FORMAT <<< }  \dimnota=2 \unoduecol=1 \fi
\magnification=\ingrandimento
%
%
%
%
\newdimen\eucolumnsize \newdimen\eudoublehsize \newdimen\eudoublevsize
\newdimen\uscolumnsize \newdimen\usdoublehsize \newdimen\usdoublevsize
\newdimen\eusinglehsize \newdimen\eusinglevsize \newdimen\ussinglehsize
\newskip\standardbaselineskip \newdimen\ussinglevsize
\newskip\reducedbaselineskip \newskip\doublebaselineskip
\eucolumnsize=12.0truecm    
\eudoublehsize=25.5truecm   
\eudoublevsize=6.7truein    
\uscolumnsize=4.4truein     
\usdoublehsize=9.4truein    
\usdoublevsize=6.8truein    
\eusinglehsize=6.5truein    
\eusinglevsize=24truecm     
\ussinglehsize=6.5truein    
\ussinglevsize=8.9truein    
\standardbaselineskip=16pt plus.2pt  
\reducedbaselineskip=14pt plus.2pt   
\doublebaselineskip=12pt plus.2pt    
%
%
\def\Portoffset{}
\def\Landoffset{\voffset=-.2truein}
\ifx\risposta\mplarisposta \def\Portoffset{\hoffset=1.8truecm} \fi
%
%
\def\Landspec{}
\tolerance=10000
\parskip=0pt plus2pt  \leftskip=0pt \rightskip=0pt
%
%
\ifx\risposta\standardrisposta \infralinea=\standardbaselineskip \fi
\ifx\risposta\reducedrisposta  \infralinea=\reducedbaselineskip \fi
\ifx\risposta\doublerisposta   \infralinea=\doublebaselineskip \fi
\ifx\risposta\mplarisposta     \infralinea=13pt \fi
\ifx\risposta\zerorisposta     \infralinea=12pt plus.2pt\fi
\ifnum\controllorisposta=0    \infralinea=\standardbaselineskip \fi
\ifx\risposta\doublerisposta   \Landoffset \else \Portoffset \fi
\ifx\risposta\doublerisposta \ifx\srisposta\cartarisposta
\tothsize=\eudoublehsize \collhsize=\eucolumnsize
\vsize=\eudoublevsize  \else  \tothsize=\usdoublehsize
\collhsize=\uscolumnsize \vsize=\usdoublevsize \fi \else
\ifx\srisposta\cartarisposta \tothsize=\eusinglehsize
\vsize=\eusinglevsize \else  \tothsize=\ussinglehsize
\vsize=\ussinglevsize \fi \collhsize=4.4truein \fi
\ifx\risposta\mplarisposta \tothsize=5.0truein
\vsize=7.8truein \collhsize=4.4truein \fi
%
%
%
%
\newcount\contaeuler \newcount\contacyrill \newcount\contaams
\font\ninerm=cmr9  \font\eightrm=cmr8  \font\sixrm=cmr6
\font\ninei=cmmi9  \font\eighti=cmmi8  \font\sixi=cmmi6
\font\ninesy=cmsy9  \font\eightsy=cmsy8  \font\sixsy=cmsy6
\font\ninebf=cmbx9  \font\eightbf=cmbx8  \font\sixbf=cmbx6
\font\ninett=cmtt9  \font\eighttt=cmtt8  \font\nineit=cmti9
\font\eightit=cmti8 \font\ninesl=cmsl9  \font\eightsl=cmsl8
\skewchar\ninei='177 \skewchar\eighti='177 \skewchar\sixi='177
\skewchar\ninesy='60 \skewchar\eightsy='60 \skewchar\sixsy='60
\hyphenchar\ninett=-1 \hyphenchar\eighttt=-1 \hyphenchar\tentt=-1
%
\font\tencmmib=cmmib10  \newfam\cmmibfam  \skewchar\tencmmib='177
\font\tencmbsy=cmbsy10  \newfam\cmbsyfam  \skewchar\tencmbsy='60
\font\tencmcsc=cmcsc10  \newfam\cmcscfam
\ifnum\ingrandimento=1095

\else

\fi

\def\ttaarr{\bf}                
\def\ppaarr{\sl}                

%
%
%
\newfam\eufmfam \newfam\msamfam \newfam\msbmfam \newfam\eufbfam
\def\Loadeulerfonts{\global\contaeuler=1 \ifx\arisposta\amsrisposta
\font\teneufm=eufm10              
\font\eighteufm=eufm8 \font\nineeufm=eufm9 \font\sixeufm=eufm6
\font\seveneufm=eufm7  \font\fiveeufm=eufm5
\font\teneufb=eufb10              
\font\eighteufb=eufb8 \font\nineeufb=eufb9 \font\sixeufb=eufb6
\font\seveneufb=eufb7  \font\fiveeufb=eufb5
\font\teneurm=eurm10              
\font\eighteurm=eurm8 \font\nineeurm=eurm9
\font\teneurb=eurb10              
\font\eighteurb=eurb8 \font\nineeurb=eurb9
\font\teneusm=eusm10              
\font\eighteusm=eusm8 \font\nineeusm=eusm9
\font\teneusb=eusb10              
\font\eighteusb=eusb8 \font\nineeusb=eusb9
\else \def\eufm{\tt} \def\eufb{\tt} \def\eurm{\tt} \def\eurb{\tt}
\def\eusm{\tt} \def\eusb{\tt}    \fi}

\def\loadamsmath{\global\contaams=1 \ifx\arisposta\amsrisposta
\font\tenmsam=msam10 \font\ninemsam=msam9 \font\eightmsam=msam8
\font\sevenmsam=msam7 \font\sixmsam=msam6 \font\fivemsam=msam5
\font\tenmsbm=msbm10 \font\ninemsbm=msbm9 \font\eightmsbm=msbm8
\font\sevenmsbm=msbm7 \font\sixmsbm=msbm6 \font\fivemsbm=msbm5
\else \def\msbm{\bf} \fi \def\Bbb{\msbm} \def\symbl{\msam} \tenpoint}
\def\loadcyrill{\global\contacyrill=1 \ifx\arisposta\amsrisposta
\font\tenwncyr=wncyr10 \font\ninewncyr=wncyr9 \font\eightwncyr=wncyr8
\font\tenwncyb=wncyr10 \font\ninewncyb=wncyr9 \font\eightwncyb=wncyr8
\font\tenwncyi=wncyr10 \font\ninewncyi=wncyr9 \font\eightwncyi=wncyr8
\else \def\cyrill{\sl} \def\cyrilb{\sl} \def\cyrili{\sl} \fi\tenpoint}
\ifx\arisposta\amsrisposta
\font\sevenex=cmex7               
\font\eightex=cmex8  \font\nineex=cmex9
\font\ninecmmib=cmmib9   \font\eightcmmib=cmmib8
\font\sevencmmib=cmmib7 \font\sixcmmib=cmmib6
\font\fivecmmib=cmmib5   \skewchar\ninecmmib='177
\skewchar\eightcmmib='177  \skewchar\sevencmmib='177
\skewchar\sixcmmib='177   \skewchar\fivecmmib='177
\font\ninecmbsy=cmbsy9    \font\eightcmbsy=cmbsy8
\font\sevencmbsy=cmbsy7  \font\sixcmbsy=cmbsy6
\font\fivecmbsy=cmbsy5   \skewchar\ninecmbsy='60
\skewchar\eightcmbsy='60  \skewchar\sevencmbsy='60
\skewchar\sixcmbsy='60    \skewchar\fivecmbsy='60
\font\ninecmcsc=cmcsc9    \font\eightcmcsc=cmcsc8     \else
\def\cmmib{\fam\cmmibfam\tencmmib}\textfont\cmmibfam=\tencmmib
\scriptfont\cmmibfam=\tencmmib \scriptscriptfont\cmmibfam=\tencmmib
\def\cmbsy{\fam\cmbsyfam\tencmbsy} \textfont\cmbsyfam=\tencmbsy
\scriptfont\cmbsyfam=\tencmbsy \scriptscriptfont\cmbsyfam=\tencmbsy
\scriptfont\cmcscfam=\tencmcsc \scriptscriptfont\cmcscfam=\tencmcsc
\def\cmcsc{\fam\cmcscfam\tencmcsc} \textfont\cmcscfam=\tencmcsc \fi
\catcode`@=11
\newskip\ttglue
\gdef\tenpoint{\def\rm{\fam0\tenrm}
  \textfont0=\tenrm \scriptfont0=\sevenrm \scriptscriptfont0=\fiverm
  \textfont1=\teni \scriptfont1=\seveni \scriptscriptfont1=\fivei
  \textfont2=\tensy \scriptfont2=\sevensy \scriptscriptfont2=\fivesy
  \textfont3=\tenex \scriptfont3=\tenex \scriptscriptfont3=\tenex
  \def\mcal{\fam2 \tensy}  \def\mmit{\fam1 \teni}
  \textfont\itfam=\tenit \def\it{\fam\itfam\tenit}
  \textfont\slfam=\tensl \def\sl{\fam\slfam\tensl}
  \textfont\ttfam=\tentt \scriptfont\ttfam=\eighttt
  \scriptscriptfont\ttfam=\eighttt  \def\tt{\fam\ttfam\tentt}
  \textfont\bffam=\tenbf \scriptfont\bffam=\sevenbf
  \scriptscriptfont\bffam=\fivebf \def\bf{\fam\bffam\tenbf}
     \ifx\arisposta\amsrisposta    \ifnum\contaeuler=1
  \textfont\eufmfam=\teneufm \scriptfont\eufmfam=\seveneufm
  \scriptscriptfont\eufmfam=\fiveeufm \def\eufm{\fam\eufmfam\teneufm}
  \textfont\eufbfam=\teneufb \scriptfont\eufbfam=\seveneufb
  \scriptscriptfont\eufbfam=\fiveeufb \def\eufb{\fam\eufbfam\teneufb}
  \def\eurm{\teneurm} \def\eurb{\teneurb} \def\eusm{\teneusm}
  \def\eusb{\teneusb}    \fi    \ifnum\contaams=1
  \textfont\msamfam=\tenmsam \scriptfont\msamfam=\sevenmsam
  \scriptscriptfont\msamfam=\fivemsam \def\msam{\fam\msamfam\tenmsam}
  \textfont\msbmfam=\tenmsbm \scriptfont\msbmfam=\sevenmsbm
  \scriptscriptfont\msbmfam=\fivemsbm \def\msbm{\fam\msbmfam\tenmsbm}
     \fi      \ifnum\contacyrill=1     \def\cyrill{\tenwncyr}
  \def\cyrilb{\tenwncyb}  \def\cyrili{\tenwncyi}         \fi
  \textfont3=\tenex \scriptfont3=\sevenex \scriptscriptfont3=\sevenex
  \def\cmmib{\fam\cmmibfam\tencmmib} \scriptfont\cmmibfam=\sevencmmib
  \textfont\cmmibfam=\tencmmib  \scriptscriptfont\cmmibfam=\fivecmmib
  \def\cmbsy{\fam\cmbsyfam\tencmbsy} \scriptfont\cmbsyfam=\sevencmbsy
  \textfont\cmbsyfam=\tencmbsy  \scriptscriptfont\cmbsyfam=\fivecmbsy
  \def\cmcsc{\fam\cmcscfam\tencmcsc} \scriptfont\cmcscfam=\eightcmcsc
  \textfont\cmcscfam=\tencmcsc \scriptscriptfont\cmcscfam=\eightcmcsc
     \fi            \tt \ttglue=.5em plus.25em minus.15em
  \normalbaselineskip=12pt
  \setbox\strutbox=\hbox{\vrule height8.5pt depth3.5pt width0pt}
  \let\sc=\eightrm \let\big=\tenbig   \normalbaselines
  \baselineskip=\infralinea  \rm}
\gdef\ninepoint{\def\rm{\fam0\ninerm}
  \textfont0=\ninerm \scriptfont0=\sixrm \scriptscriptfont0=\fiverm
  \textfont1=\ninei \scriptfont1=\sixi \scriptscriptfont1=\fivei
  \textfont2=\ninesy \scriptfont2=\sixsy \scriptscriptfont2=\fivesy
  \textfont3=\tenex \scriptfont3=\tenex \scriptscriptfont3=\tenex
  \def\mcal{\fam2 \ninesy}  \def\mmit{\fam1 \ninei}
  \textfont\itfam=\nineit \def\it{\fam\itfam\nineit}
  \textfont\slfam=\ninesl \def\sl{\fam\slfam\ninesl}
  \textfont\ttfam=\ninett \scriptfont\ttfam=\eighttt
  \scriptscriptfont\ttfam=\eighttt \def\tt{\fam\ttfam\ninett}
  \textfont\bffam=\ninebf \scriptfont\bffam=\sixbf
  \scriptscriptfont\bffam=\fivebf \def\bf{\fam\bffam\ninebf}
     \ifx\arisposta\amsrisposta  \ifnum\contaeuler=1
  \textfont\eufmfam=\nineeufm \scriptfont\eufmfam=\sixeufm
  \scriptscriptfont\eufmfam=\fiveeufm \def\eufm{\fam\eufmfam\nineeufm}
  \textfont\eufbfam=\nineeufb \scriptfont\eufbfam=\sixeufb
  \scriptscriptfont\eufbfam=\fiveeufb \def\eufb{\fam\eufbfam\nineeufb}
  \def\eurm{\nineeurm} \def\eurb{\nineeurb} \def\eusm{\nineeusm}
  \def\eusb{\nineeusb}     \fi   \ifnum\contaams=1
  \textfont\msamfam=\ninemsam \scriptfont\msamfam=\sixmsam
  \scriptscriptfont\msamfam=\fivemsam \def\msam{\fam\msamfam\ninemsam}
  \textfont\msbmfam=\ninemsbm \scriptfont\msbmfam=\sixmsbm
  \scriptscriptfont\msbmfam=\fivemsbm \def\msbm{\fam\msbmfam\ninemsbm}
     \fi       \ifnum\contacyrill=1     \def\cyrill{\ninewncyr}
  \def\cyrilb{\ninewncyb}  \def\cyrili{\ninewncyi}         \fi
  \textfont3=\nineex \scriptfont3=\sevenex \scriptscriptfont3=\sevenex
  \def\cmmib{\fam\cmmibfam\ninecmmib}  \textfont\cmmibfam=\ninecmmib
  \scriptfont\cmmibfam=\sixcmmib \scriptscriptfont\cmmibfam=\fivecmmib
  \def\cmbsy{\fam\cmbsyfam\ninecmbsy}  \textfont\cmbsyfam=\ninecmbsy
  \scriptfont\cmbsyfam=\sixcmbsy \scriptscriptfont\cmbsyfam=\fivecmbsy
  \def\cmcsc{\fam\cmcscfam\ninecmcsc} \scriptfont\cmcscfam=\eightcmcsc
  \textfont\cmcscfam=\ninecmcsc \scriptscriptfont\cmcscfam=\eightcmcsc
     \fi            \tt \ttglue=.5em plus.25em minus.15em
  \normalbaselineskip=11pt
  \setbox\strutbox=\hbox{\vrule height8pt depth3pt width0pt}
  \let\sc=\sevenrm \let\big=\ninebig \normalbaselines\rm}
\gdef\eightpoint{\def\rm{\fam0\eightrm}
  \textfont0=\eightrm \scriptfont0=\sixrm \scriptscriptfont0=\fiverm
  \textfont1=\eighti \scriptfont1=\sixi \scriptscriptfont1=\fivei
  \textfont2=\eightsy \scriptfont2=\sixsy \scriptscriptfont2=\fivesy
  \textfont3=\tenex \scriptfont3=\tenex \scriptscriptfont3=\tenex
  \def\mcal{\fam2 \eightsy}  \def\mmit{\fam1 \eighti}
  \textfont\itfam=\eightit \def\it{\fam\itfam\eightit}
  \textfont\slfam=\eightsl \def\sl{\fam\slfam\eightsl}
  \textfont\ttfam=\eighttt \scriptfont\ttfam=\eighttt
  \scriptscriptfont\ttfam=\eighttt \def\tt{\fam\ttfam\eighttt}
  \textfont\bffam=\eightbf \scriptfont\bffam=\sixbf
  \scriptscriptfont\bffam=\fivebf \def\bf{\fam\bffam\eightbf}
     \ifx\arisposta\amsrisposta   \ifnum\contaeuler=1
  \textfont\eufmfam=\eighteufm \scriptfont\eufmfam=\sixeufm
  \scriptscriptfont\eufmfam=\fiveeufm \def\eufm{\fam\eufmfam\eighteufm}
  \textfont\eufbfam=\eighteufb \scriptfont\eufbfam=\sixeufb
  \scriptscriptfont\eufbfam=\fiveeufb \def\eufb{\fam\eufbfam\eighteufb}
  \def\eurm{\eighteurm} \def\eurb{\eighteurb} \def\eusm{\eighteusm}
  \def\eusb{\eighteusb}       \fi    \ifnum\contaams=1
  \textfont\msamfam=\eightmsam \scriptfont\msamfam=\sixmsam
  \scriptscriptfont\msamfam=\fivemsam \def\msam{\fam\msamfam\eightmsam}
  \textfont\msbmfam=\eightmsbm \scriptfont\msbmfam=\sixmsbm
  \scriptscriptfont\msbmfam=\fivemsbm \def\msbm{\fam\msbmfam\eightmsbm}
     \fi       \ifnum\contacyrill=1     \def\cyrill{\eightwncyr}
  \def\cyrilb{\eightwncyb}  \def\cyrili{\eightwncyi}         \fi
  \textfont3=\eightex \scriptfont3=\sevenex \scriptscriptfont3=\sevenex
  \def\cmmib{\fam\cmmibfam\eightcmmib}  \textfont\cmmibfam=\eightcmmib
  \scriptfont\cmmibfam=\sixcmmib \scriptscriptfont\cmmibfam=\fivecmmib
  \def\cmbsy{\fam\cmbsyfam\eightcmbsy}  \textfont\cmbsyfam=\eightcmbsy
  \scriptfont\cmbsyfam=\sixcmbsy \scriptscriptfont\cmbsyfam=\fivecmbsy
  \def\cmcsc{\fam\cmcscfam\eightcmcsc} \scriptfont\cmcscfam=\eightcmcsc
  \textfont\cmcscfam=\eightcmcsc \scriptscriptfont\cmcscfam=\eightcmcsc
     \fi             \tt \ttglue=.5em plus.25em minus.15em
  \normalbaselineskip=9pt
  \setbox\strutbox=\hbox{\vrule height7pt depth2pt width0pt}
  \let\sc=\sixrm \let\big=\eightbig \normalbaselines\rm }
\gdef\tenbig#1{{\hbox{$\left#1\vbox to8.5pt{}\right.\n@space$}}}
\gdef\ninebig#1{{\hbox{$\textfont0=\tenrm\textfont2=\tensy
   \left#1\vbox to7.25pt{}\right.\n@space$}}}
\gdef\eightbig#1{{\hbox{$\textfont0=\ninerm\textfont2=\ninesy
   \left#1\vbox to6.5pt{}\right.\n@space$}}}
\def\alternativefont#1#2{\ifx\arisposta\amsrisposta \relax \else
\xdef#1{#2} \fi}
\global\contaeuler=0 \global\contacyrill=0 \global\contaams=0
%
%
%
%
\newbox\fotlinebb \newbox\hedlinebb \newbox\leftcolumn
\gdef\makeheadline{\vbox to 0pt{\vskip-22.5pt
     \fullline{\vbox to8.5pt{}\the\headline}\vss}\nointerlineskip}
\gdef\makehedlinebb{\vbox to 0pt{\vskip-22.5pt
     \fullline{\vbox to8.5pt{}\copy\hedlinebb\hfil
     \line{\hfill\the\headline\hfill}}\vss} \nointerlineskip}
\gdef\makefootline{\baselineskip=24pt \fullline{\the\footline}}
\gdef\makefotlinebb{\baselineskip=24pt
    \fullline{\copy\fotlinebb\hfil\line{\hfill\the\footline\hfill}}}
\gdef\doubleformat{\shipout\vbox{\Landspec\makehedlinebb
     \fullline{\box\leftcolumn\hfil\columnbox}\makefotlinebb}
     \advancepageno}
\gdef\columnbox{\leftline{\pagebody}}
\gdef\line#1{\hbox to\hsize{\hskip\leftskip#1\hskip\rightskip}}
\gdef\fullline#1{\hbox to\fullhsize{\hskip\leftskip{#1}%
\hskip\rightskip}}
\gdef\footnote#1{\let\@sf=\empty
         \ifhmode\edef\#sf{\spacefactor=\the\spacefactor}\/\fi
         #1\@sf\vfootnote{#1}}
\gdef\vfootnote#1{\insert\footins\bgroup
         \ifnum\dimnota=1  \eightpoint\fi
         \ifnum\dimnota=2  \ninepoint\fi
         \ifnum\dimnota=0  \tenpoint\fi
         \interlinepenalty=\interfootnotelinepenalty
         \splittopskip=\ht\strutbox
         \splitmaxdepth=\dp\strutbox \floatingpenalty=20000
         \leftskip=\oldssposta \rightskip=\olddsposta
         \spaceskip=0pt \xspaceskip=0pt
         \ifnum\sinnota=0   \textindent{#1}\fi
         \ifnum\sinnota=1   \item{#1}\fi
         \footstrut\futurelet\next\fo@t}
\gdef\fo@t{\ifcat\bgroup\noexpand\next \let\next\f@@t
             \else\let\next\f@t\fi \next}
\gdef\f@@t{\bgroup\aftergroup\@foot\let\next}
\gdef\f@t#1{#1\@foot} \gdef\@foot{\strut\egroup}
\gdef\footstrut{\vbox to\splittopskip{}}
\skip\footins=\bigskipamount
\count\footins=1000  \dimen\footins=8in
\catcode`@=12
\tenpoint
\ifnum\unoduecol=1 \hsize=\tothsize   \fullhsize=\tothsize \fi
\ifnum\unoduecol=2 \hsize=\collhsize  \fullhsize=\tothsize \fi
\global\let\lrcol=L      \ifnum\unoduecol=1
\output{\plainoutput{\ifnum\tipbnota=2 \clearnmbnota\fi}} \fi
\ifnum\unoduecol=2 \output{\if L\lrcol
     \global\setbox\leftcolumn=\columnbox
     \global\setbox\fotlinebb=\line{\hfill\the\footline\hfill}
     \global\setbox\hedlinebb=\line{\hfill\the\headline\hfill}
     \advancepageno  \global\let\lrcol=R
     \else  \doubleformat \global\let\lrcol=L \fi
     \ifnum\outputpenalty>-20000 \else\dosupereject\fi
     \ifnum\tipbnota=2\clearnmbnota\fi }\fi
\def\ifdoublepage{\ifnum\unoduecol=2 }
\gdef\yespagenumbers{\footline={\hss\tenrm\folio\hss}}
\gdef\ciao{ \ifnum\fdefcontre=1 \endfdef\fi
     \par\vfill\supereject \ifnum\unoduecol=2
     \if R\lrcol  \headline={}\nopagenumbers\null\vfill\eject
     \fi\fi \end}

\newskip\olddsposta \newskip\oldssposta
\global\oldssposta=\leftskip \global\olddsposta=\rightskip

\def\filldots{\leaders\hbox to 1em{\hss.\hss}\hfill}
\def\inquadrb#1 {\vbox {\hrule  \hbox{\vrule \vbox {\vskip .2cm
    \hbox {\ #1\ } \vskip .2cm } \vrule  }  \hrule} }
 \def\newline{\hfil\break}
\def\jump{\vskip\baselineskip} \newskip\iinnffrr
\def\sjump{\iinnffrr=\baselineskip
          \divide\iinnffrr by 2 \vskip\iinnffrr}
\def\bjump{\vskip\baselineskip \vskip\baselineskip}
\newcount\nmbnota  \def\clearnmbnota{\global\nmbnota=0}
\newcount\tipbnota \def\letterfootnote{\global\tipbnota=1}

\def\note#1{\global\advance\nmbnota by 1 \ifnum\tipbnota=1
    \footnote{$^{\rm\nttlett}$}{#1} \else {\ifnum\tipbnota=2
    \footnote{$^{\nttsymb}$}{#1}
    \else\footnote{$^{\the\nmbnota}$}{#1}\fi}\fi}
\def\nttlett{\ifcase\nmbnota \or a\or b\or c\or d\or e\or f\or
g\or h\or i\or j\or k\or l\or m\or n\or o\or p\or q\or r\or
s\or t\or u\or v\or w\or y\or x\or z\fi}
\def\nttsymb{\ifcase\nmbnota \or\dag\or\sharp\or\ddag\or\star\or
\natural\or\flat\or\clubsuit\or\diamondsuit\or\heartsuit
\or\spadesuit\fi}   \clearnmbnota
\def\numberfootnote{\global\tipbnota=0} \numberfootnote
\def\setnote#1{\expandafter\xdef\csname#1\endcsname{
\ifnum\tipbnota=1 {\rm\nttlett} \else {\ifnum\tipbnota=2
{\nttsymb} \else \the\nmbnota\fi}\fi} }
\newcount\nbmfig  \def\clearnbmfig{\global\nbmfig=0}
\gdef\figure{\global\advance\nbmfig by 1
      {\rm fig. \the\nbmfig}}   \clearnbmfig
\def\setfig#1{\expandafter\xdef\csname#1\endcsname{fig. \the\nbmfig}}

\newcount\frmcount \def\clearfrmcount{\global\frmcount=0}
\def\numero{\global\advance\frmcount by 1   \ifnum\indappcount=0
  {\ifnum\cpcount <1 {\hbox{\rm (\the\frmcount )}}  \else
  {\hbox{\rm (\the\cpcount .\the\frmcount )}} \fi}  \else
  {\hbox{\rm (\applett .\the\frmcount )}} \fi}
\def\nameformula#1{\global\advance\frmcount by 1%
\ifnum\draftnum=0  {\ifnum\indappcount=0%
{\ifnum\cpcount<1\xdef\spzzttrra{(\the\frmcount )}%
\else\xdef\spzzttrra{(\the\cpcount .\the\frmcount )}\fi}%
\else\xdef\spzzttrra{(\applett .\the\frmcount )}\fi}%
\else\xdef\spzzttrra{(#1)}\fi%
\expandafter\xdef\csname#1\endcsname{\spzzttrra}
\eqno \hbox{\rm\spzzttrra} $$}
\def\nfr{\nameformula}    
\def\nameali#1{\global\advance\frmcount by 1%
\ifnum\draftnum=0  {\ifnum\indappcount=0%
{\ifnum\cpcount<1\xdef\spzzttrra{(\the\frmcount )}%
\else\xdef\spzzttrra{(\the\cpcount .\the\frmcount )}\fi}%
\else\xdef\spzzttrra{(\applett .\the\frmcount )}\fi}%
\else\xdef\spzzttrra{(#1)}\fi%
\expandafter\xdef\csname#1\endcsname{\spzzttrra}
  \hbox{\rm\spzzttrra} }      \clearfrmcount
\newcount\cpcount \def\clearcpcount{\global\cpcount=0}
\newcount\subcpcount \def\clearsubcpcount{\global\subcpcount=0}
\newcount\appcount \def\clearappcount{\global\appcount=0}
\newcount\indappcount \def\clearindappcount{\indappcount=0}
\newcount\sottoparcount 

\def\applett{\ifcase\appcount  \or {A}\or {B}\or {C}\or
{D}\or {E}\or {F}\or {G}\or {H}\or {I}\or {J}\or {K}\or {L}\or
{M}\or {N}\or {O}\or {P}\or {Q}\or {R}\or {S}\or {T}\or {U}\or
{V}\or {W}\or {X}\or {Y}\or {Z}\fi    \ifnum\appcount<0
\immediate\write16 {Panda ERROR - Appendix: counter "appcount"
out of range}\fi  \ifnum\appcount>26  \immediate\write16 {Panda
ERROR - Appendix: counter "appcount" out of range}\fi}
\clearappcount  \clearindappcount \newcount\connttrre
\def\clearconnttrre{\global\connttrre=0} \newcount\countref
\def\clearcountref{\global\countref=0} \clearcountref
\def\chapter#1{\global\advance\cpcount by 1 \clearfrmcount
                 \goodbreak\null\vbox{\sjump\nobreak
                 \clearsubcpcount\clearindappcount
                 \itemitem{\ttaarr\the\cpcount .\qquad}{\ttaarr #1}
                 \par\nobreak\sjump}\nobreak}
\def\section#1{\global\advance\subcpcount by 1 \goodbreak\null
               \vbox{\sjump\nobreak\ifnum\indappcount=0
                 {\ifnum\cpcount=0 {\itemitem{\ppaarr
               .\the\subcpcount\quad\enskip\ }{\ppaarr #1}\par} \else
                 {\itemitem{\ppaarr\the\cpcount .\the\subcpcount\quad
                  \enskip\ }{\ppaarr #1} \par}  \fi}
                \else{\itemitem{\ppaarr\applett .\the\subcpcount\quad
                 \enskip\ }{\ppaarr #1}\par}\fi\nobreak\jump}\nobreak}
\clearsubcpcount
\def\appendix#1{\global\advance\appcount by 1 \clearfrmcount
                  \goodbreak\null\vbox{\jump\nobreak
                  \global\advance\indappcount by 1 \clearsubcpcount
          \itemitem{ }{\hskip-40pt\ttaarr Appendix\ \applett :\ #1}
             \nobreak\jump\sjump}\nobreak}
\clearappcount \clearindappcount
\def\references{\goodbreak\null\vbox{\jump\nobreak
   \itemitem{}{\ttaarr References} \nobreak\jump\sjump}\nobreak}

\clearcpcount\clearcountref

\def\setchap#1{\ifnum\indappcount=0{\ifnum\subcpcount=0%
\xdef\spzzttrra{\the\cpcount}%
\else\xdef\spzzttrra{\the\cpcount .\the\subcpcount}\fi}
\else{\ifnum\subcpcount=0 \xdef\spzzttrra{\applett}%
\else\xdef\spzzttrra{\applett .\the\subcpcount}\fi}\fi
\expandafter\xdef\csname#1\endcsname{\spzzttrra}}
\newcount\draftnum \newcount\ppora   \newcount\ppminuti
\global\ppora=\time   \global\ppminuti=\time
\global\divide\ppora by 60  \draftnum=\ppora
\multiply\draftnum by 60    \global\advance\ppminuti by -\draftnum
\def\droggi{\number\day /\number\month /\number\year\ \the\ppora
:\the\ppminuti}     \global\draftnum=0
\def\draftcomment#1{\ifnum\draftnum=0 \relax \else
{\ {\bf ***}\ #1\ {\bf ***}\ }\fi} 
%
%
\catcode`@=11
\gdef\Ref#1{\expandafter\ifx\csname @rrxx@#1\endcsname\relax%
{\global\advance\countref by 1    \ifnum\countref>200
\immediate\write16 {Panda ERROR - Ref: maximum number of references
exceeded}  \expandafter\xdef\csname @rrxx@#1\endcsname{0}\else
\expandafter\xdef\csname @rrxx@#1\endcsname{\the\countref}\fi}\fi
\ifnum\draftnum=0 \csname @rrxx@#1\endcsname \else#1\fi}
\gdef\beginref{\ifnum\draftnum=0  \gdef\Rref{\fairef}
\gdef\endref{\scriviref} \else\relax\fi
\ifx\risposta\mplarisposta \ninepoint \fi
\parskip 2pt plus.2pt \baselineskip=12pt}
\def\Reflab#1{[#1]} \gdef\Rref#1#2{\item{\Reflab{#1}}{#2}}
\gdef\endref{\relax}  \newcount\conttemp
\gdef\fairef#1#2{\expandafter\ifx\csname @rrxx@#1\endcsname\relax
{\global\conttemp=0 \immediate\write16 {Panda ERROR - Ref: reference
[#1] undefined}} \else
{\global\conttemp=\csname @rrxx@#1\endcsname } \fi
\global\advance\conttemp by 50  \global\setbox\conttemp=\hbox{#2} }
\gdef\scriviref{\clearconnttrre\conttemp=50
\loop\ifnum\connttrre<\countref \advance\conttemp by 1
\advance\connttrre by 1
\item{\Reflab{\the\connttrre}}{\unhcopy\conttemp} \repeat}
\clearcountref \clearconnttrre
\catcode`@=12
\ifx\risposta\mplarisposta \def\Reflab#1{#1.} \letterfootnote \fi

\def\slashchar#1{\setbox0=\hbox{$#1$} \dimen0=\wd0
     \setbox1=\hbox{/} \dimen1=\wd1 \ifdim\dimen0>\dimen1
      \rlap{\hbox to \dimen0{\hfil/\hfil}} #1 \else
      \rlap{\hbox to \dimen1{\hfil$#1$\hfil}} / \fi}
\ifx\oldchi\undefined \let\oldchi=\chi
  \def\cchi{{\raise 1pt\hbox{$\oldchi$}}} \let\chi=\cchi \fi
  
  \def\alfa{\alpha} 

\def\frac#1#2{{\textstyle{#1 \over #2}}}

\def\half{\ifinner {\scriptstyle {1 \over 2}}\else {1 \over 2} \fi}

\def\simge{\rlap{\raise 2pt \hbox{$>$}}{\lower 2pt \hbox{$\sim$}}}
\def\simle{\rlap{\raise 2pt \hbox{$<$}}{\lower 2pt \hbox{$\sim$}}}

\def\vbig#1#2{{\vbigd@men=#2\divide\vbigd@men by 2%
\hbox{$\left#1\vbox to \vbigd@men{}\right.\n@space$}}}

%
%
\newcount\fdefcontre \newcount\fdefcount \newcount\indcount
\newread\filefdef  \newread\fileftmp  \newwrite\filefdef
\newwrite\fileftmp     \def\strip#1*.A {#1}
\def\futuredef#1{\beginfdef
\expandafter\ifx\csname#1\endcsname\relax%
{\immediate\write\fileftmp {#1*.A}
\immediate\write16 {Panda Warning - fdef: macro "#1" on page
\the\pageno \space undefined}
\ifnum\draftnum=0 \expandafter\xdef\csname#1\endcsname{(?)}
\else \expandafter\xdef\csname#1\endcsname{(#1)} \fi
\global\advance\fdefcount by 1}\fi   \csname#1\endcsname}

\def\beginfdef{\ifnum\fdefcontre=0
\immediate\openin\filefdef \jobname.fdef
\immediate\openout\fileftmp \jobname.ftmp
\global\fdefcontre=1  \ifeof\filefdef \immediate\write16 {Panda
WARNING - fdef: file \jobname.fdef not found, run TeX again}
\else \immediate\read\filefdef to\spzzttrra
\global\advance\fdefcount by \spzzttrra
\indcount=0      \loop\ifnum\indcount<\fdefcount
\advance\indcount by 1   \immediate\read\filefdef to\spezttrra
\immediate\read\filefdef to\sppzttrra
\edef\spzzttrra{\expandafter\strip\spezttrra}
\immediate\write\fileftmp {\spzzttrra *.A}
\expandafter\xdef\csname\spzzttrra\endcsname{\sppzttrra}
\repeat \fi \immediate\closein\filefdef \fi}
\def\endfdef{\immediate\closeout\fileftmp   \ifnum\fdefcount>0
\immediate\openin\fileftmp \jobname.ftmp
\immediate\openout\filefdef \jobname.fdef
\immediate\write\filefdef {\the\fdefcount}   \indcount=0
\loop\ifnum\indcount<\fdefcount    \advance\indcount by 1
\immediate\read\fileftmp to\spezttrra
\edef\spzzttrra{\expandafter\strip\spezttrra}
\immediate\write\filefdef{\spzzttrra *.A}
\edef\spezttrra{\string{\csname\spzzttrra\endcsname\string}}
\iwritel\filefdef{\spezttrra}
\repeat  \immediate\closein\fileftmp \immediate\closeout\filefdef
\immediate\write16 {Panda Warning - fdef: Label(s) may have changed,
re-run TeX to get them right}\fi}
\def\iwritel#1#2{\newlinechar=-1
{\newlinechar=`\ \immediate\write#1{#2}}\newlinechar=-1}
\global\fdefcontre=0 \global\fdefcount=0 \global\indcount=0
%
%
\null
%
%
%
%